# Convergent and divergent connectivity patterns of the arcuate fasciculus in macaques and humans

Jiahao Huang[1,†], Ruifeng Li[1,†], Wenwen Yu[2], Anan Li[3,4], Xiangning Li[4], Mingchao Yan[5], Lei Xie[1], Qingrun Zeng[1], Xueyan Jia[3], Shuxin Wang[3], Ronghui Ju[6,7], Feng Chen[8], Qingming Luo[4], Hui Gong[3], Andrew Zalesky[9], Xiaoquan Yang[3,4,*], Yuanjing Feng[1,*], and Zheng Wang[4,10,*]

[†]These authors contributed equally to this work.

Author affiliations:

[1]Institute of Advanced Technology, College of Information Engineering, Zhejiang University of Technology, Hangzhou, China

[2]Institute of Science and Technology for Brain-inspired Intelligence, Fudan University, Shanghai 200433, China

[3]HUST-Suzhou Institute for Brainsmatics, JITRI, Suzhou 215123, China

[4]State Key Laboratory of Digital Medical Engineering, Key Laboratory of Biomedical Engineering of Hainan Province, School of Biomedical Engineering, Hainan University, Hainan, China

[5]Institute of Neuroscience, State Key Laboratory of Neuroscience, Center for Excellence in Brain Science and Intelligence Technology, Chinese Academy of Sciences, Shanghai, China

[6]Department of Radiology, The People's Hospital of China Medical University and the People's Hospital of Liaoning Province, Shenyang 110016, China

[7]Liaoning Provincial Key Laboratory of Neurointerventional Therapy and Biomaterials Research and Development, Shenyang, China

[8]Department of Radiology, Hainan General Hospital (Hainan Affiliated Hospital of Hainan Medical University), Haikou 570311, Hainan, China

[9]Systems Lab, Department of Psychiatry, Melbourne Medical School, The University of Melbourne, Melbourne, Victoria, Australia

[10]School of Psychological and Cognitive Sciences; Beijing Key Laboratory of Behavior and Mental Health; IDG/McGovern Institute for Brain Research; Peking-Tsinghua Center for Life Sciences, Peking University, Beijing, China

**Correspondence to:**

1.* Xiaoquan Yang, PhD.

HUST-Suzhou Institute for Brainsmatics, JITRI, Suzhou 215123, China

Email: xqyang@mail.hust.edu.cn

2.* Yuanjing Feng, PhD.

Institute of Information Processing and Automation, College of Information Engineering, Zhejiang University of Technology, Hangzhou 310023, China

Email: fyjing@zjut.edu.cn

3.* Zheng Wang, PhD.

School of Psychological and Cognitive Sciences; Beijing Key Laboratory of Behavior and Mental Health; IDG/McGovern Institute for Brain Research; Peking-Tsinghua Center for Life Sciences, Peking University, Beijing, China

Email: zheng.wang@pku.edu.cn

**Running title**: AF connectome in primates

# Significance

Language is the foundation of human communication and emotional expression, and its evolution is crucial for understanding human behavior and social interaction. The arcuate fasciculus (AF) is a key neural fiber supporting language, but the exact trajectory of the AF in non-human primates and its differences from humans remain unclear. In this study, we reconstructed the AF pathway in macaques using fluorescent Micro-Optical Sectioning Tomography and ultrahigh-field diffusion imaging, and compared it with humans. We found similar connectivity between the auditory cortex and temporoparietal region in both species, while the human AF shows greater expansion in the prefrontal cortex, supramarginal gyrus, and middle temporal gyrus. Our data and methods provide tools for studying language network evolution and understanding language origins.

# Abstract

The organization and connectivity of the arcuate fasciculus (AF) in nonhuman primates remain contentious, especially concerning how its anatomy diverges from that of humans. Here, we combined cross-scale single-neuron tracing - using viral-based genetic labeling and fluorescence micro-optical sectioning tomography in macaques (n = 4; age 3 - 11 years) - with whole-brain tractography from 11.7T diffusion MRI. Complemented by spectral embedding analysis of 7.0T MRI in humans, we performed a comparative connectomic analysis of the AF across species. We demonstrate that the macaque AF originates in the temporal-parietal cortex, traverses the auditory cortex and parietal operculum, and projects into prefrontal regions. In contrast, the human AF exhibits greater expansion into the middle temporal gyrus and stronger prefrontal and parietal operculum connectivity - divergences quantified by Kullback-Leibler analysis that likely underpin the evolutionary specialization of human language networks. These differences underscore the critical role of AF expansion and differentiation in the evolution of human language capabilities.

# Introduction

Language, a uniquely human capability, serves as the foundation for cognition, culture, and social interaction (1-5), profoundly reliant on the fronto-temporal network and its dorsal-ventral language streams (6-12). The arcuate fasciculus (AF), the dorsal pathway's core white matter tract, connects auditory and ventrolateral prefrontal cortices (vlPFC), supporting complex syntax, phonology and auditory processing (9, 14-16). However, debates persist regarding AF's projections to divisions of temporal regions and its distinctness from the superior longitudinal fasciculus (SLF) due to overlapping termination (17-19). While comparative neuroanatomy has identified homologs of the AF in nonhuman primates (15, 19, 20), the divergent connectivity patterns that may underlie human-unique linguistic capacities remain unresolved. Evolutionary studies have shown that the AF exhibits specific expansion and unique anatomical features in the auditory cortex and temporal speech perception regions, and thus the core language network anchored on the AF is often regarded as an independent, dedicated language system (21). However, this view has been increasingly challenged, as studies relying on fixed "language localizer" tasks often overlook the contributions of other complex brain subnetworks to language function, highlighting the need for a broader exploration of the language network (22). Further elucidating the neuroanatomical organization of the AF is not only critical for understanding the evolution of language, but also paves the way for uncovering the biological mechanisms underlying language functional networks (6, 21, 22).

Diffusion MRI (dMRI) tractography has mapped interspecies AF variations (23, 24), with multi-modal approaches (e.g., diffusion spectrum imaging and isotope tracing studies) validating Sylvian fissure connectivity (25). Recent work using functional seeding further confirm the presence of AF's auditory-anchored pathways in macaques, chimpanzees and humans (15). Yet the human AF demonstrates expanded temporal (middle/inferior temporal gyri) and reinforced frontoparietal connectivity—divergences hypothesized to facilitate hierarchical syntax and semantic integration (15, 20, 25, 26). Nevertheless, conflicting evidence persists regarding AF's projection to basal temporal cortex, and its topographical overlap with SLF pathways, limiting evolutionary interpretations (19). Crucially, whether these differences reflect quantitative scaling or qualitative reorganization remains untested, in part due to the lack of cross-species connectivity mapping at cellular resolution. Advanced optical microscopic imaging technology combined with fluorescent sparse labeling, high-precision slicing, continuous signal detection, and single cell tracing has enabled the mesoscale reconstruction of brain-wide, cell-type specific projections in rodents and macaques (27-29), which also provides a ground-truth framework for validating dMRI reconstructions. Integrating these approaches could resolve long-standing debates - does the human AF incorporate novel cortical targets absent in primates, or does it amplify ancestral circuits? Such insights would refine models of how connective innovations underpin language emergence and identify homologous substrates for translational studies.

Here, we unify fluorescent Micro-Optical Sectioning Tomography (fMOST) and 11.7T diffusion tractography to establish the first single-neuron-resolved AF connectome in macaques (n = 4), coupled with 7.0T human dMRI from the Human Connectome Project (30). Our multi-scale, multi-modal pipeline: maps vlPFC-to-temporal AF projections at cellular scale (0.65 × 0.65 × 3.00 μm$^3$) using anterograde/retrograde viral tracers (**Figures. 1A and 2A-D**); reconstructs whole-brain AF connectome with neuronally constrained tractography in a 3D space based on ex vivo dMRI data of the monkey brain (500 × 500 × 500 μm$^3$) (**Figure. 1B**); and quantifies cross-species divergence and similarities via spectral embedding of cortical connectivity and Kullback-Leibler analysis (**Figure. 1C**). This integrative approach not only clarifies AF's evolutionary trajectory but also provides a pathoanatomical framework for future investigation of language disorders - linking degraded AF connectivity profiles in aphasia to evolutionarily derived circuit vulnerabilities.

# Results

## Single-neuron reconstruction and visualization of AF in macaque monkeys

Guided by anatomical MRI images of the monkey brain based on our previous work (31-33), we injected both anterograde and retrograde viruses: AAV2/9-CaMKIIa-Tau-GFP and retroAAV-CaMKIIa-mCherry to label the vlPFC neurons and to trace bi-directional axonal projections with two different colors (green and red) from fMOST imaging (**Figures. 2A and 3A-F**). The injection site in the vlPFC, validated by fMOST images, including areas 44, 45a/b, 46v/f, and 12r/l, was precisely located in cortical gray matter (**Supplementary Figure. 1**). The pipeline for imaging the macaque brain with fMOST included a multi-step process, and more details are provided in the Methods and **Figures. 2A-D**. To acquire a detailed account of the brain-wide vlPFC connectome, we analyzed its connectivity profile based on the statistical analysis of the fluorescence signals distributed in the macaque brain atlas using the fMOST data. Results (**Figure. 3G**) show that the fluorescence signals were predominantly concentrated in the frontal lobe (65.66%), and the temporal lobe (17.90%), occipital lobe (7.70%), parietal lobe (5.26%), limbic regions (2.00%), and insular regions (1.48%). Within the frontal lobe, signals were primarily found in areas 12l, 44, and 45a, whereas signals in the temporal lobe were mainly concentrated in some visual and auditory processing areas (more details are provided in **Supplementary Material and video. 1**).

Importantly, we have reconstructed individual AF neurons with a resolution of 0.65 × 0.65 × 3.00 μm$^3$ based on the fMOST data (more details are described in the Methods), which allowed to identify the terminal areas of axonal projections in the targeted brain regions (**Figures. 2C-D**). We found that those reconstructed neurons primarily terminated in areas Tpt, TPO, rostromedial belt region of the auditory cortex (RM), and superior temporal sulcus (sts) part of the temporal pole (**Figures. 3H-O**), essentially forming the dorsal and ventral parts, some of which generally exceeds 20,000 μm (**Figure. 3I**). Two representative dorsal AF neurons had endpoints located in the Tpt and TPO (**Figure. 3J and Supplementary video. 2**). These neurons arched around the caudal part of the Sylvian fissure's depth, initially passing through the white matter in the ventral and dorsal parts of the parietal operculum. They then traveled beyond the extreme and external capsules to enter area 45 of the vlPFC (neuron 1 and 2 in **Figures. 3J-K and 4A**). At the temporal-parietal junction, we observed neuronal projections in the dorsal white matter of the 7op area (**Figure. 3K**). These distinct projection paths at this junction result in a visual difference in the "bow" structure between neuron 1 and neuron 2 of the AF pathway. Interestingly, we observed other AF neuron projections that curved downward around the lateral fissure, passed through Tpt and terminated at the junction of TPO and TAa (neuron 3 in **Figures. 3M and 4A**). **Figure. 3O** illustrates the trajectory detail of this neuron, including the morphology of the soma and dendrites, a 3D view of axon bifurcation, and the corresponding information shown in the original fMOST images (**Supplementary video. 3**). Additionally, the bifurcated axons passed above the caudal medial belt (CM) and the parietal lobe (neuron 3 and neuron 4 in **Figure. 4A**), forming a complete neural loop between the frontal and temporal lobes. It may indicate a potential information exchange pathway between the dorsal part of the AF pathway and the auditory cortex (13, 15).

# Whole-brain reconstruction and visualization of the AF connectome in macaques and humans

To characterize the connection profiles of the AF connectome, we derived the whole-brain tractography for AF tracts using 11.7T ex vivo dMRI data of the macaque brain. First, we co-registered the fMOST images and dMRI images into a common space, mapped the trajectories of all traced neurons onto dMRI images (**Figure. 4A**) and then applied as anatomical constraints to refine the template for macaque AF tractography (**Figure. 4B**) (24). Specifically, we designated the white matter regions with dense AF neuron projections as tracking seeds (**Figures. 4C-E**) and made the refined macaque AF tractography templates publicly available online (https://github.com/HuangJH9721/Cross-species_correlation.git). Finally, we estimated the AF streamlines using probabilistic tractography based on a spherical deconvolution model (34), as demonstrated in **Figure. 4F**, and the corresponding 2D mapping results in **Figure. 4G**. Evidently, the AF fiber tracts traversed the internal white matter of the parietal area 7op (**Figure. 4G**). Nevertheless, we applied the same tractography algorithm to estimate the main fiber bundles of the dorsal language pathway (AF, SLF1-3), ventral language pathway (MdLF, IFOF, ILF, and UF), and other major fiber bundles in both monkey and human brains (**Supplementary Figure 2**).

To quantify the spatial correspondence between fMOST-and dMRI-derived AF pathways (**Figure. 4H**), we calculated the Szymkiewicz-Simpson overlap coefficient in a common 3D space (**Figure. 4I**). Overall, results obtained from two modalities had a good correspondence in 3D space, with an average R of 0.4219 (STD=0.246, 95% CI = 0.364-0.479, the highest overlap = 0.88). Further statistical analysis was conducted by dividing the AF pathways (**Figure. 4I**) into anterior (42-65 slices), middle (66-89 slices), and posterior (90-113 slices) segments. We found larger consistency in the middle (average = 0.553, STD = 0.172, 95% CI = 0.481-0.626) and posterior segments (average = 0.577, STD = 0.148, 95% CI = 0.515-0.640), while relatively lower in the anterior segment (average = 0.135, STD = 0.08, 95% CI = 0.101-0.168). Nevertheless, both methods showed the presence of the AF pathway in the 7op dorsal white matter at the coronal slices (**Figure. 4J**), and in the auditory regions at the sagittal slices (**Figure. 4I**).

We then generated average structural connectivity maps of the AF for both species (11), illustrating the connection between frontal cortical regions and other brain areas via the AF (**Supplementary Figure 3**). In humans, the AF demonstrates stronger connectivity between frontal areas 44, 45, the inferior frontal junction, the auditory cortex in the temporal lobe, and areas 39 and 40 in the inferior parietal lobule. Similarly, in macaques, the AF shows strong connectivity between homologous regions, including areas 44, 45, and the auditory cortex. However, compared to macaques, the human AF connects with a larger number of brain regions, particularly in the temporal lobe, extending to areas such as the middle and inferior temporal gyri (**Supplementary Figure 3**).

## Characterization of AF connectivity profiles across both species

We next constructed a connectivity blueprint, which has been demonstrated as a powerful approach to comparative anatomy (24, 35, 36), to characterize the connectivity principles and unique specializations of AF pathways across both species. The connectivity blueprints were built as the dot product of the whole-brain connectivity matrix (cortex × whole-brain) and the tract-map matrix (whole-brain × tracts), in which the corresponding tracts were used as landmarks to define a common connectivity space for interspecies brain comparisons (**Figure. 5A**). A column of the connectivity blueprint represents the cortical territories of a tract, with results of AF tracts across different brains shown in **Figure. 5B**. The cortical terminations of human AF exhibited varying degrees of expansion in the frontal, parietal, and temporal lobes relative to macaques (**Figure. 5B**). Apparently, the dorsal termination regions of monkey AF were prominently concentrated around the posterior end of the lateral fissure (ls) and mostly did not cross the sts (**Figure. 5B**), while its human homolog clearly extended downward to the posterior part of the superior temporal gyrus and the tail of middle and inferior temporal gyri (**Figure. 5B**).

Meanwhile, a row of the connectivity blueprint describes the pattern of how a cortical vertex connects to the considered tracts, which can be used to represent the probabilistic distributions of different tracts at the same vertex by normalization (36). As such, we normalized each row of the blueprint matrix (vertex-wise) and used Brainnetome (37) and D99 atlases (38) to parcellate these vertices, and to compute the average probability distribution for each region in both species (**Supplementary Figure 4**). The results of Mann-Whitney U Test (**Figure. 5C**) showed that average probability (average ± STD) of the dorsal pathway (area 44: $0.62 ± 0.21$; area 45: $0.33 ± 0.16$) was significantly higher (area 44: $p= 2.23 × 10^{-20}$, area 45: $p= 2.51 × 10^{-4}$) than the ventral pathway (area 44: $0.11 ± 0.08$; area 45: $0.20 ± 0.09$) in human areas 44 and 45. By contrast, there was no significant difference in the monkey brain between the dorsal (area 44: $0.46 ± 0.12$; area 45: $0.36 ± 0.18$) and ventral (area 44: $0.24 ± 0.04$; area 45: $0.26 ± 0.11$) pathways. In human area 12l, the probability of the ventral pathway was significantly higher than the dorsal part (dorsal: $0.09 ± 0.06$, ventral: $0.57 ± 0.10$, $p= 6.80 × 10^{-8}$), which was different from the monkey brain (dorsal: $0.42 ± 0.16$, ventral: $0.30 ± 0.08$). Regarding the contribution of cortical territories to the AF pathways, we found that in the human frontal lobe higher probabilities of AF tracts were mainly located in areas inferior frontal junction ($0.30 ± 0.06$) and 44 (opercular area 44, $0.16 ± 0.06$; dorsal area 44, $0.30 ± 0.06$; ventral area 44, $0.17 ± 0.07$), in contrast to the monkey brain such as areas 44 ($0.22 ±$

0.08), 45 (45a, 0.19 ± 0.07; 45b, 0.22 ± 0.07), and ventral 8A (0.26 ± 0.14) (**Figure. 5D**). In the temporal lobe, differences in the proportion of cortical territories contribution to the AF pathway was also evident between human areas dorsolateral area 37 (0.46 ± 0.13), rostro posterior sts (0.36 ± 0.10) and caudo posterior sts (0.41 ± 0.09) and monkey areas Tpt (0.37 ± 0.06) and CM (0.26 ± 0.03).

Furthermore, we calculated the symmetric KL divergence to measure statistical similarity between two species at the vertex and region levels (**Supplementary Figure 5**) and mapped onto the brain surface for visualization (**Figure. 6A**) (36). It clearly indicated that species-related cortical differences were prominent in the inferior frontal cortex, inferior parietal lobe, and temporal regions (mainly the sts) of the human brain, whereas the caudal part of superior temporal areas in the monkey brain closely resembled the human homolog (**Figure. 6B**). In addition, we set an empirically defined threshold (tract contribution > 0.10) of the joint similarity matrix to focus on AF-connected brain regions (**Figure. 6C and Supplementary Figure 6**). Among these AF-connected regions, the frontal lobe exhibited significantly larger average KL divergence (1.89 ± 1.19) than the temporal (1.35 ± 0.62, FDR corrected $p < 0.05$) and parietal lobes (1.13 ± 0.53, FDR corrected $p < 0.01$) (**Figure. 6D**).

## Joint-embedding for AF-connected regions across species

We aligned AF connectivity blueprints of both monkey and human to a spectral embedding space based on a joint similarity matrix (39, 40), in which regions with similar connectivity patterns tend to cluster together. It showed that AF-connected frontal and temporal regions of two species formed distinct groups (**Figure. 6E**). And we calculated the Euclidean distance between clusters within one species to quantify the similarities and observed a significant difference in the frontal lobe between two species ($p = 1.25 \times 10^{-5}$, **Figure. 6F**), indicating substantial differentiation of human AF-connected frontal areas relative to the monkey homolog. We also calculated the Euclidean distance across two species and found that the distances between homologous regions within temporal lobe (0.037 ± 0.018) are significantly lower than that between regions in the frontal lobe (0.074 ± 0.047, FDR corrected $p < 0.01$) (**Figure. 6G**). The results reveal that the temporal regions of the AF connectome in macaques are more similar to their homologous regions in the human brain compared to the frontal regions. Moreover, upon a closer look (**Figure. 6H**), AF-connected areas 44 (0.06 ± 0.04) and 45 (0.08 ± 0.05) showed larger Euclidean distance and were mapped onto the human brain surface to demonstrate species difference within the frontal lobe. But in the temporal lobes, regions like CL (0.02 ± 0.01), CM (0.04 ± 0.02) and Tpt (0.05 ± 0.01) strongly resembled the human homolog in cortical connection and expansion.

# Materials and methods

## Animals and ethics statement

All experimental procedures for nonhuman primate research in this study were approved by the Animal Care and Use Committee of Center for Excellence in Brain Science and Intelligence Technology, Chinese Academy of Sciences, and conformed to the National Institutes of Health guidelines for the humane care and use of laboratory animals. Four macaque monkeys (3 males 1 female; aged 3-11 year; body weight 3.5-12.2 kg) were used in this project (**Supplementary Table. 7**). All subjects underwent ex-vivo ultrahigh field dMRI scanning and fMOST imaging. The animal handling procedures are described in detail in our previous work (31, 32, 41-43) and briefly summarized here.

## MRI-guided virus injection

We used AAV2/9 (AAV2/9-CaMKIIα-Tau-GFP, titer: 8.47 × 1013 vg/mL; AAV2/9-CaMKIIa-mCherry, titer: 1.6 × 1013 vg/mL) based on our previous experience (29, 31). The recombinant AAV2/9 virus included an CaMKIIα promoter to control the expression of the reporter gene. All four macaques received injections of the virus AAV2/9-CaMKIIa-Tau-GFP (anterograde) in the right vlPFC (encompassing 44, 45a/b, 12r/l, and 46f/v), and one of them received an injection of the virus retroAAV-CaMKIIa-mCherry (retrograde). To precisely target brain regions in individual subjects, we performed in-vivo MRI scanning on monkeys using a 3T scanner (Siemens Tim Trio, Erlangen, Germany) under general anesthesia to obtain T1-weighted images, which were then used to guide virus injection. Anesthesia was induced by an intramuscular injection of ketamine (10 mg/kg), followed by the maintenance of deep anesthesia with isoflurane (1.5–3%), while vital physiological signals were continuously monitored during the MRI scanning. Anatomical scans were acquired with an MPRAGE sequence using the following parameters: TR = 2300ms, TE = 2.8ms, TI = 1100ms, spatial resolution 0.5 mm isotropic. Targeting the regions of interest in the macaque brain involved using the symmetric normalization (SyN) algorithm to register the three-dimensional digital atlas of Saleem and Logothetis to individual T1-weighted images. Subsequently, the position of the vlPFC relative to the stereotaxic space was calculated. All virus injection procedures were performed under strict aseptic conditions, as described in detail in our previous work (29, 31).

## Sample preparation

Depending on the expression time of individual virus (**Supplementary Table. 7**), the monkey was transcardially perfused with 4 L of 4°C 0.01 mol/L phosphate-buffered saline (PBS) and 1 L of 4% paraformaldehyde (PFA), and anesthetized with pentobarbital (45 mg/kg, i.m.). The deployment of 4 % PFA has been detailed in our previous studies (29) and is briefly summarized here. A 4% PFA solution was prepared by dissolving 40 g of paraformaldehyde powder in 1 L of 0.01 mol/L PBS solution. Following perfusion, the macaque brains were extracted and stored in the 4% PFA solution for subsequent immunofluorescence staining and imaging. The fixed intact macaque brain was immersed in a Poly-N-acryloyl glycinamide (PNAGA) hydrogel solution at 4°C. Then, the water-soluble azo initiator VA-044 was added to initiate the polymerization of N-acryloyl glycinamide (NAGA) (29). After achieving complete tissue permeation, a 4-hour polymerization process at 40°C was conducted. Finally, the macaque brain-hydrogel hybrid, at room temperature, was prepared for subsequent micro-optical sectioning tomography.

## Macaque ex-vivo dMRI data

We collected ex-vivo dMRI data from four macaques with an 11.7T MR scanner. Prior to conducting MRI scans, the brain samples underwent several preprocessing steps to enhance their suitability for imaging. These steps included immersion of the specimens in a gadolinium-based contrast agent, which served to shorten the T1 relaxation time and improve data acquisition efficiency. Additionally, FOMBLIN perfluoropolyether was employed to prevent susceptibility-induced artifacts. To further optimize the samples for MRI scanning, they were subjected to a 24-hour degassing process under controlled pressure conditions, effectively eliminating air bubbles (31). MR images of the macaque brains were acquired using an 11.7T animal MRI system (31). We employed a 3D diffusion-weighted spin–echo pulse sequence with single-line readout, with which key parameters were set as follows: TR/ TE = 82/22.19 ms, FOV = 64×58 mm², isotropic spatial resolution = 0.5 mm, 60 noncollinear diffusion gradient directions with b = 4000 s/mm² ($\Delta/\delta$ = 15/2.8 ms, maximal gradient strength = 746.56 mT/m) and 5 non-diffusion b0 images. The MRI images of each monkey brain were registered to the macaque D99 template using the SyN in ANTs (44).

## Human in-vivo dMRI data

Human dMRI data of the HCP dataset (30) were acquired on a 7.0T MRI scanner (syngo MR B17, Siemens Magnetom Investigational Device, Erlangen, Germany), using a diffusion-weighted spin–echo echo planar imaging sequence, with which key imaging parameters were set as follows: TR = 7000 ms; TE = 71.2 ms; isotropic spatial resolution = 1.05 mm; FOV = 210 × 210 mm²; flip angle = 90°; 128 diffusion weighting directions with two shells at b=1000, 2000 s/mm² and 15 non-diffusion b0 images. The human MRI images were also registered to the MNI152 template using the SyN in ANTs (44).

## Fluorescence Micro-Optical Sectioning Tomography

The fMOST system was customized to accommodate the macaque monkey brain for dual-channel imaging (29, 45). Two lasers (Cobolt, Solna, Sweden) of different wavelengths (488 nm and 561 nm) were combined using a dichroic mirror and expanded with a beam expander (f = 7.5 mm and 250 mm). The beams were then shaped by a cylindrical lens and projected onto the sample using a concave lens (f = 125 mm) and the objective lens. The fluorescence collected by the objective was filtered and then received by the CMOS camera. Two CMOS cameras captured green and red fluorescence images. The sectioning unit for high-quality brain slices consisted of a motor driver, a double-leaf spring vibration mechanism, and a 57 mm carbide blade with a 25° bevel angle. Optimized parameters include an 87 Hz vibration frequency, 1 mm amplitude, and 0.5 mm/s speed, minimizing perpendicular movement to 2 μm. Sectioning and imaging are synchronized with the stage.

The imaging process was conducted as follows: initially, the contour of the entire sample was captured. Subsequently, the sample was then sectioned to an appropriate thickness using a vibratome, and corresponding images were acquired. This slicing and imaging process was repeated iteratively until the entire sample was fully imaged. **(Supplementary video. 1)**. To reduce image blurring caused by surface fluctuations, the initial imaging focal plane was typically positioned 10 μm deeper than the slicing plane. Three-dimensional images were captured using a 10x objective, with a voxel size of 0.65 × 0.65 × 3.00 μm³. Red fluorescence images, captured through the red channel, provided structural information about the cells on the surface of the macaque brain, with somas labeled using mCherry. Green fluorescence images, captured through the green channel, revealed neurons that were anterogradely labeled with GFP.

## Single-neuron tracing of monkey AF

The image preprocessing for the two detection channels included lateral illumination correction via polynomial fitting to determine correction coefficients, followed by stitching the stripes of each imaging plane based on spatial orientation and overlap. The stitched stripes were combined to form a single image for each imaging plane. These images were then stored in a 16-bit TIFF format. The parallel image preprocessing process was conducted by C++ and the Intel MPI Library (v3.2.2.006, USA). Preprocessing the images of one macaque brain (voxel size: $0.65 \times 0.65 \times 3.00$ μm$^3$) on a 280-core, 2.4 GHz computing server was completed within 40 days. The image size for each macaque brain is detailed in **Supplementary Table. 7.** To manage the TB-sized data on a single workstation, we converted the original TIFF format data to the TDat format (**Figure. 2**) (46).

We reconstructed the macaque brain-wide tracings of long-range axons in 3D using a human-computer interaction approach. The process involved confirming the spatial position of the region of interest on the Z-axis maximum projection map (a portion of the macaque data is available online at https://github.com/HuangJH9721/Cross-species_correlation.git). Data blocks along the axons and dendrites were then continuously read into the software (**Supplementary video. 3**). The start and end points of each signal segment within a neuron were manually verified for each TDat block, and the software automatically calculated the labeled path of the neuron between these points (**Figure. 2**). This process was repeated until the complete trajectory of a single neuron was reconstructed in 3D (**Supplementary video. 2**). The macaque fMOST data were visualized using Amira software (v5.4.1, FEI, Mérignac Cedex, France) to generate the images. The outline of the whole macaque brain was manually delineated and segmented using Amira's Segmentation Editor module with low-resolution data ($300 \times 300 \times 300$ μm$^3$).

## Brain-wide tractography of AF

The constrained spherical deconvolution model, a mathematical model proposed for multiple fiber orientation estimation within each voxel (47), was adopted to estimate the orientation of fibers in the macaque and human brains based on dMRI data using MRtrix 3 (https://www.mrtrix.org/) (34). To ensure the comparability of tractography results between macaque and human brains, we employed a probabilistic tractography method using MRtrix 3's 'iFod2' algorithm (34). This method has previously demonstrated superior efficacy in tracking white matter fibers in our research (48, 49). For the macaque brain, the tractography template was adapted by modifying the existing macaque template available in FSL (50). In the case of the human brain, the necessary mask, seed image, and region of interest (ROI) for tractography were derived using TractSeg (51). The parameters utilized for the probabilistic tractography were meticulously chosen: curvature threshold = 20°; seed cutoff = 0.05; stop cutoff = 0.05; number of streamlines = 20000; step size (for humans) = 0.05; step size (for macaques) = 0.02; maximum length = 250 mm; and minimum length = 10 mm. Using this approach, we performed tractography of several major white matter tracts in both human and macaque brains, including the AF, inferior fronto-occipital fasciculus (IFOF), inferior longitudinal fasciculus (ILF), middle longitudinal fasciculus (MdLF), superior longitudinal fasciculus (SLF1, SLF2, SLF3), uncinate fasciculus (UF), anterior thalamic radiation (ATR), optic radiation (OR), corticospinal tract (CST), and posterior thalamic radiation (PTR). To generate track density maps, we utilized MRtrix 3's 'tckmap' function, which calculates the number of streamlines passing through different voxels and normalizes the result, providing a comprehensive visualization of the tractography data.

## Cross-scale comparison of the monkey AF pathway by fMOST and dMRI

To facilitate a direct comparison between the diffusion-derived AF tract and fMOST-derived axonal fibers, we first generated track density images of the AF tract using the tckmap tool, which were subsequently co-registered to the template space using the SyN. For the creation of dMRI-derived fiber tract volumes, we applied a threshold of 0.5 to the density maps, converting any voxel below this threshold to zero, thereby producing a binarized mask of the AF tract. Concurrently, fluorescence images obtained from fMOST data were downsampled to a resolution of $30 \times 30 \times 100$ μm$^3$. Manual annotation of AF axons was performed across 2708 coronal slices of these images. To ensure accuracy, signals from surrounding neurons were validated within 3D blocks of $650 \times 650 \times 650$ μm$^3$ using the Gtree software. The annotated 2D outlines of the AF pathway were then aggregated across slices to generate 3D renderings of the vlPFC projection pathways using Amira software (version 5.4.1, FEI, Mérignac Cedex,

France), ultimately resulting in a binarized mask of the vlPFC projection. The Szymkiewicz-Simpson overlap coefficient was adopted to quantify the spatial overlap between the two sets of data, defined as the size of their union over the size of the smaller set:

$$overlap(AFn, AFt) = \frac{|AFn \cap AFt|}{\min(|AFn|, |AFt|)} \tag{1}$$

The Szymkiewicz-Simpson overlap coefficient ranges from 0 (indicating no overlap) to 1 (indicating complete overlap), in which $AF_n$ represented the binary maps derived from fMOST images, while $AF_t$ corresponded to the maps generated through diffusion tractography.

## Building connectivity blueprints

Based on tractography results, we employed an established approach known as "connectivity blueprint" to extract connectivity profiles for both humans and macaques (24, 35). The connectivity blueprint is represented as a matrix (cortex × tracts), where the tract dimensions are consistent across both species. Each row of the connectivity blueprint represents the probability distribution of major tracts across the cortex, while each column details the cortical terminations of each tract. To begin, we constructed a whole-brain connectivity matrix (cortex × whole-brain) by initiating streamlines from the cortical white-gray matter boundary using probabilistic tractography and quantifying visitations between seeds and white matter voxels throughout the brain. Subsequently, we generated a tract-map matrix (whole-brain × tracts) by mapping the tractography results of various tracts into a one-dimensional form, recording connections between tracts and white matter voxels. The final connectivity blueprints (cortex × tracts) were then derived by calculating the product of the whole-brain connectivity matrix and the tract-map matrix.

Using this methodology, we constructed connectivity blueprints for both the human and macaque brains. Additionally, group-averaged blueprints were derived by averaging the connectivity blueprints generated for each subject within each dataset. Specifically, the average connectivity blueprints (group-wise) for humans were generated using 7.0T dMRI data from 20 adult subjects in the HCP cohort, while the macaque blueprints were derived from dMRI data obtained from four ex vivo macaque brains.

# Cross-species comparison of the AF pathway in macaque and human brains

Each column of the normalized group-averaged blueprint can be interpreted as a probability distribution within a defined sample space, allowing for the application of Kullback-Leibler (KL) divergence to compare these distributions across different species. Consequently, we employed the symmetric KL divergence as a measure of similarity between these distributions:

$$D_{ik} = \sum_{k} M_{ik} \log_2 \frac{M_{ik}}{H_{jk}} + \sum_{k} H_{jk} \log_2 \frac{H_{jk}}{M_{ik}} \qquad (2)$$

where H and M denote the group-averaged blueprints for the human and macaque brains, respectively. The term $M_{ik}$ and $M_{jk}$ represent the likelihood of a connection from vertices i and j on the macaque cortex to tract k, while $H_{ik}$ and $H_{jk}$ denote the corresponding likelihoods for the human cortex. The KL divergence was computed between each column of a k × i connectivity profile and a k × j connectivity profile, yielding an i × j KL divergence matrix. This KL divergence matrix was then used to quantify the (dis-)similarity of cross-species vertices on a vertex-wise basis. To generate parcellated KL divergence matrices (region-wise), we segmented the vertex-wise KL divergence matrix (20002 human vertices × 20002 macaque vertices) along the columns and rows using the corresponding cortical parcellation atlas (37, 41). And the median KL divergence for each parcel was calculated to represent the (dis-)similarities among parcels. This process resulted in the formation of a region-wise KL divergence matrix (120 human cortical regions × 184 macaque cortical regions).

We constructed a symmetrical joint KL divergence matrix (39, 40), which concatenates within- and cross-species region-wise KL divergence matrices: $W_{Human}$, $W_{Macaque}$, $W_{Human\_to\_Macaque}$, and $W_{Macaque\_to\_Human}$ as follows:

$$W_{Joint} = [W_{Human}, W_{Human\_to\_Macaque}; W_{Macaque}, W_{Macaque\_to\_Human}] \qquad (3)$$

Thus, a negative transformation was applied to the JKL matrix, followed by the application of an exponential function to derive the similarity matrix. This transformation effectively converts the dissimilarity measure into a similarity measure, facilitating spectral embedding analysis. Using a spectral embedding method, we then projected the similarity matrix into a 2D space. This 2D representation intuitively highlighted both the commonalities and disparities in cross-species cortical connectivity patterns. The spectral embedding method groups parcels with similar connectivity profiles together within the projected space. The similarity between brain regions in macaque and human brains was quantitatively assessed by measuring the Euclidean distance between corresponding parcels in the 2D embedding space. Subsequently, connectivity embedding and pairwise distance calculations were performed on these cortical regions to further compare the variability in the connectivity profiles of the AF-connected cortices.

## Discussion

Previous studies have shown that the macaque brain processes both spatial and non-spatial auditory information through the interplay between the dorsal and ventral auditory pathways, along with the non-primary auditory cortex in the caudal and rostral regions (13). Furthermore, recent evidence identifies the AF as a crucial tract within the dorsal auditory pathway, demonstrating its connections with the auditory cortex (15). In this study, using single-neuron mapping with fMOST, we observed long-range axonal fiber connections between the vlPFC and the caudal superior temporal lobe. This provides direct evidence of the spatial trajectory of the monkey AF pathway, complementing findings from previous MRI-based studies (13, 15). Our results indicate that the dorsal auditory pathway in the macaque brain, predominantly involving the AF, originates from the vlPFC, extends to the Tpt and anchors into the auditory cortex (**Figures. 3H-O and 4A**). Within the frontal lobe, isotopic techniques have shown that the macaque AF primarily projects to the dorsal premotor area 6, dorsal area 8A, and dorsal area 9/46 (25). However, single-neuron tracing reveals direct connections from the boundary between areas 45 and area 12l to Tpt, providing anatomical evidence that contrasts with findings from isotopic-based findings (**Figure. 4A**). The ongoing controversies about AF and SLF connectivity largely arise from their functional and spatial similarities, making them difficult to distinguish (18). Intriguingly, our results reveal that some AF and SLF neurons shared a common soma, forming an integrated frontal-temporal neuronal circuit (**Figure. 3M**). Notably, the visual distinctions in the arcuate structures of macaque AF neurons may be attributed to varying trajectories through the dorsal or internal white matter of the parietal lobe, particularly near the temporal lobe terminations (**Figure. 3K**). These findings unravel new anatomical insights into macaque AF, laying the groundwork for cross-species connectivity comparisons.

Findings from ex-vivo high resolution dMRI indicate that the AF primarily terminates in the vlPFC and the caudal region of the superior temporal lobe in the macaque brain, aligning closely with outcomes from neuronal tracing studies (**Figure. 4**). The tractography-derived findings of macaque brains, informed by single-neuron tracing, further enable to elucidate the whole-brain characteristics of the AF connectome. Investigations into the primate auditory system have revealed that both the auditory belt and the temporal lobe maintain direct or indirect connections with prefrontal regions (13), specifically area 8A and dorsal area 46d. Notably, area 8A, recognized as a crucial hub for the interaction of visual and auditory information, along with several regions in the auditory belt (including CM and CL), demonstrates significant contributions from the AF in their connectivity patterns (**Figure. 5D**). These findings underscore the integral role of AF connectivity within the auditory-supported fronto-temporal functional network.

Our cross-species comparisons reveal striking similarities in the connectivity profiles of the AF within the macaque auditory cortex compared to the human brain. However, there are notable differences in the connectivity characteristics of the frontal lobe, particularly regarding the expansions observed in the parietal and middle temporal lobes in humans, which are absent in macaques. A key distinction is found in the vlPFC, which shows marked differences between the two species, likely attributable to the essential role of Broca's area in facilitating motor language functions (18, 35). Previous studies have asserted that auditory language processing in macaques predominantly relies on the ventral auditory pathway (13, 15); however, our findings suggest that the dorsal auditory pathway, significantly influenced by the AF homologue, also plays a crucial role (13, 15). Research on chimpanzees further indicates that the model of human auditory functions may have originated in early non-human primates, subsequently evolving over time (16, 23, 35). To investigate whether the extension of the human AF into the temporal lobe contributes to semantic comprehension, future studies should focus on the inferior and middle temporal gyri instead of the posterior superior temporal gyrus that connects to the auditory cortex. Our results affirm the predominant role of AF connectivity in auditory language functions and pave the way for further research into the interspecies variations of AF connectomes.

Several practical limitations should be considered when interpreting our results. First, achieving sparse viral labeling remains a significant challenge; dense and mixed signals at the injection site complicate the observation of nearby axonal projections. Future research will involve viral injections into the macaque auditory cortex to investigate long-distance projections to the Wernicke homolog area and to analyze whole-brain projection patterns. Second, the evolution and differentiation of AF connectivity are ongoing processes (19). Future studies should examine AF connectivity variations across a broader spectrum of non-human primates, including both Old and New World monkeys, to enhance our understanding of the evolutionary development of the AF as a controlling bundle for language functions. Lastly, integrating recent gene expression data from transcriptome maps with our findings could offer deeper insights into the relationship between structural connectivity and gene expression (32, 52). This integration would further elucidate the mechanisms underlying AF connectivity in the evolution of language functions.

## Data availability

The resource presented here includes (l) ex vivo 11.7T dMRI data at 500 μm isotropic resolution from macaque monkeys, (2) microscopy data at 200 μm isotropic resolution from ex vivo macaque brain, (3) WM fiber pathway reconstructions, (4) macaque AF dMRI-tractography template generated in this study, (5) the connectivity blueprints of two species. The raw data and all other features of the resource can be downloaded at https://github.com/HuangJH9721/Cross-species_correlation.git. The data volume and 3D-reconstructed neuron from fMOST can be downloaded at the website (https://github.com/HuangJH9721/Cross-species_correlation.git).

Then, the Cynomolgus macaque template (Cyno162) is available at https://doi.org/10.1093/cercor/bhaa229. The human brain atlas data is available at https://doi.org/10.1093/cercor/bhw157. The human diffusion MRI datasets is available at https://www.humanconnectome.org.

## Acknowledgements

We thank members of the HUST and IDG/McGovern laboratory for discussions and comments on the manuscript. The research was supported by STI-2030-Major Projects (No. 2021ZD0204000, No. 2021ZD0200104), Zhejiang Province Science and Technology Innovation Leading Talent Program (No. 2021R52004) and the National Natural Science Foundation (No. 82151303, 61976190, 62403428 and U23A20334).

# Funding

The research was supported by STI-2030-Major Projects (No. 2021ZD0204000, No. 2021ZD0200104), Zhejiang Province Science and Technology Innovation Leading Talent Program (No. 2021R52004) and the National Natural Science Foundation (No. 82151303, 61976190, 62403428 and U23A20334).

# Competing interests

The authors report no competing interests.

# Supplementary material

(1) Supplementary Figures 1-6 with their legends and the material describing the experimental subjects and fMOST imaging are included in the Supplementary Figures and Material.

(2) Supplementary Videos 1-3 showing the details of the fMOST-derived AF pathway in macaque brains

# Figure legends

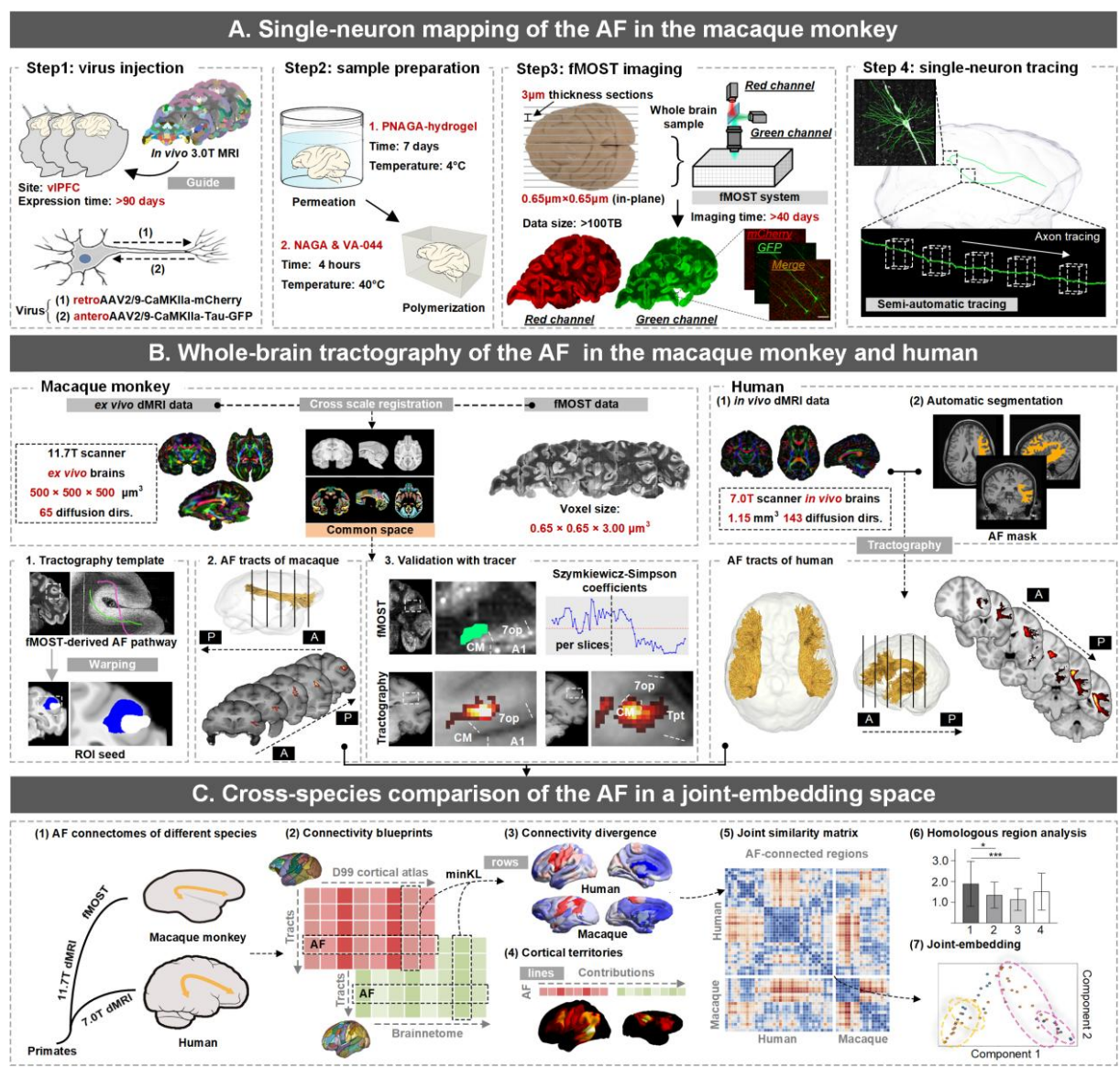

**Figure 1 A flowchart diagram for single-neuron, whole-brain mapping and interspecies comparison of the AF pathway.** (A) Viral-based genetic tracing of the AF in macaque monkeys involved stereotaxic injection of AAV vectors into the vlPFC, guided by T1-weighted MRI. After viral expression, ex-vivo macaque brains were immersed in PNAGA hydrogel for 7 days at 20°C, followed by a 4-hour VA-044 polymerization at 40°C. The embedded brains were sectioned and imaged using the fMOST system, achieving high-resolution images (0.65 × 0.65 × 3.00 μm$^3$). A semi-automatic method was used to reconstruct and visualize AF neurons at the single-neuron level. (B) Whole brain tractography of the AF in macaque monkeys was performed using ex vivo dMRI data from an 11.7T MRI scanner. Reconstructed AF neurons

provided anatomical priors to refine the AF tractography template. The AF connectome was mapped using probabilistic tractography, and dMRI-derived AF tracts were validated against fMOST tracing results. **(C)** Cross-species comparisons of the AF connectome involved computing connectivity profiles of AF-connected brain areas using dMRI-derived data. Statistical analyses were conducted on cortical territories, connectivity features, and homologous cortical similarities across species. Connectivity differences were quantified using the KL divergence matrix, and spectral embedding analysis enabled interspecies comparison in a joint-embedding space.

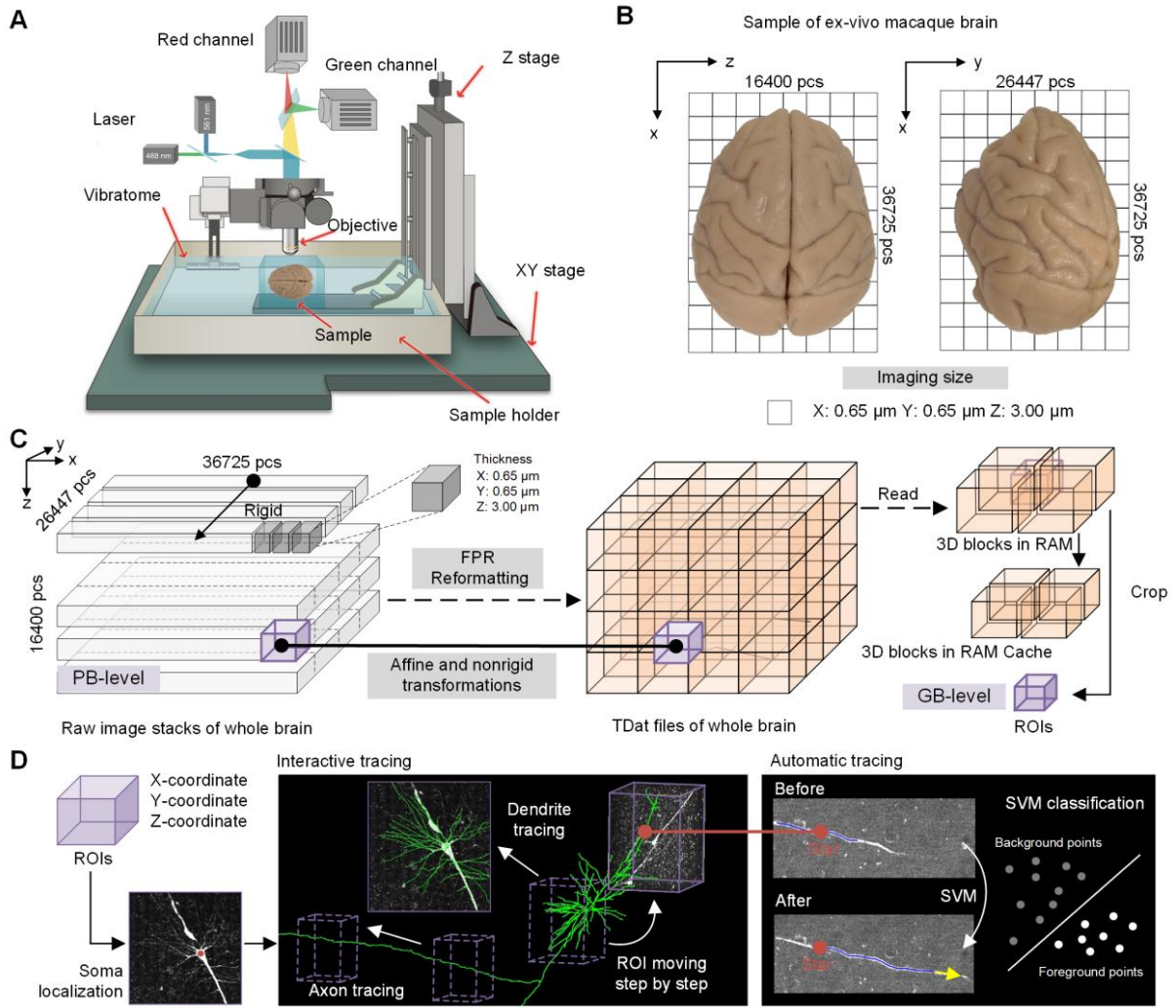

**Figure 2 Pipeline for fMOST imaging and neuronal tracing in the macaque brain. (A)** The fMOST system operates by embedding the hydrogel-fixed brain within a sample holder mounted on a 3D-stage. The sample surface is sectioned using a vibratome. High-speed data acquisition is facilitated by a two-channel line-scan confocal microscope equipped with an sCMOS camera. **(B)** An ex-vivo macaque brain sample was imaged with a raw resolution of 0.65 × 0.65 × 3.00 μm$^3$, generating imaging data with dimensions of X: 36,725 pcs, Y: 26,447 pcs, Z: 16,400 pcs. **(C)** The reformatting and data access for TDat involves the Fine-Grained Parallel Reformatting method. Data retrieval occurs at varying levels depending on the size of the region of interest (ROI). To calculate the ROI, the relevant 3D data blocks (represented in purple) are loaded into RAM, aligned according to their spatial positions, and extraneous data are removed to extract the precise ROI data. **(D)** The semi-automatic workflow for single-neuron tracing begins with identifying the soma's position by inputting the ROI coordinates into the GTree software, followed by sequentially reading ROIs along the neuron's path.

Automatic tracing of single neurons within each ROI is achieved using machine learning classifiers.

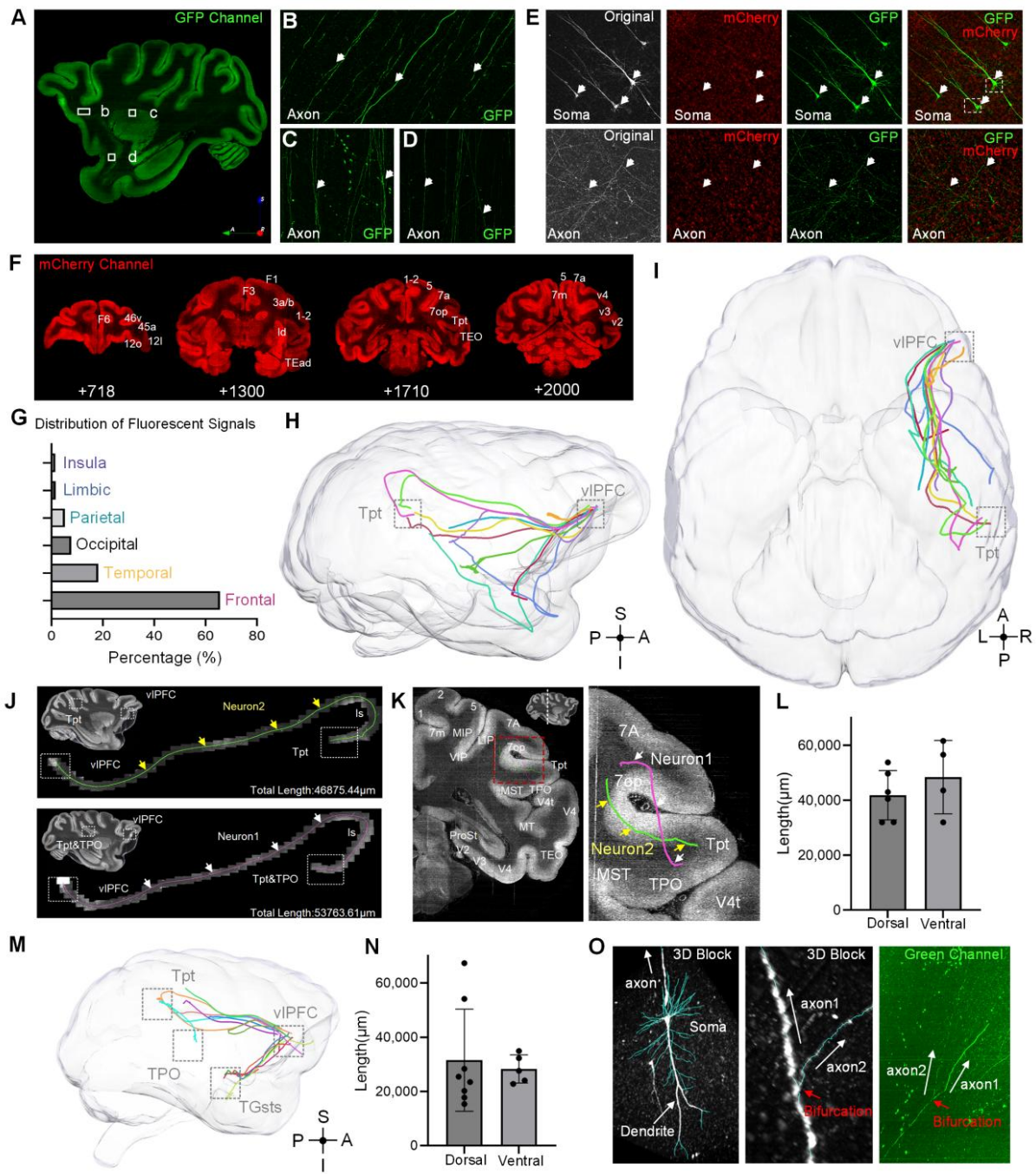

**Figure 3 Single-neuron mapping of the monkey AF pathway. (A)** Representative sagittal plane image from the GFP channel of macaque fMOST images. **(B-D)** Magnified views of the boxed regions in **(A)** reveal the morphology of GFP-labeled axons: **(B)** located between areas 45b and 13l, **(C)** within the superior region of the Striatum, and **(D)** situated between the Claustrum and areas Ial, TGa. (E) Dual-channel images displaying somas and axons near the injection site, captured in both mCherry and GFP channels. **(F)** Coronal plane images from the mCherry channel, spanning slices 718 to 2000 (rostral to the prefrontal cortex). **(G)** Brain-wide distribution of vlPFC axonal projections based on fluorescence signal statistics. **(H)** Eleven

reconstructed neurons from vlPFC projecting into the right hemisphere via dorsal and ventral pathways in a macaque brain. **(I)** Top-down view of the reconstructed neurons in **(H)**. **(J)** Two representative AF neurons, shown through a series of overlapping raw image stacks, projected from the vlPFC to the Tpt and TPO regions. **(K)** Coronal plane depiction of the fMOST image, showing the temporal-parietal area traversed by the two neurons in **(J)**. **(L)** Bar plots illustrating individual results and mean fiber length for dorsal and ventral pathways of neurons shown in **(H)** and **(I)**. **(M)** Thirteen reconstructed neurons from the vlPFC projecting into the temporal lobe via dorsal and ventral pathways in another macaque brain. **(N)** Bar plots showing the individual results and mean fiber length for dorsal and ventral pathways of the neurons displayed in **(M)**. **(O)** Images of the soma and axon bifurcation of neuron 4 (Figures 3M and 4A) in the raw image stacks and the GFP channel.

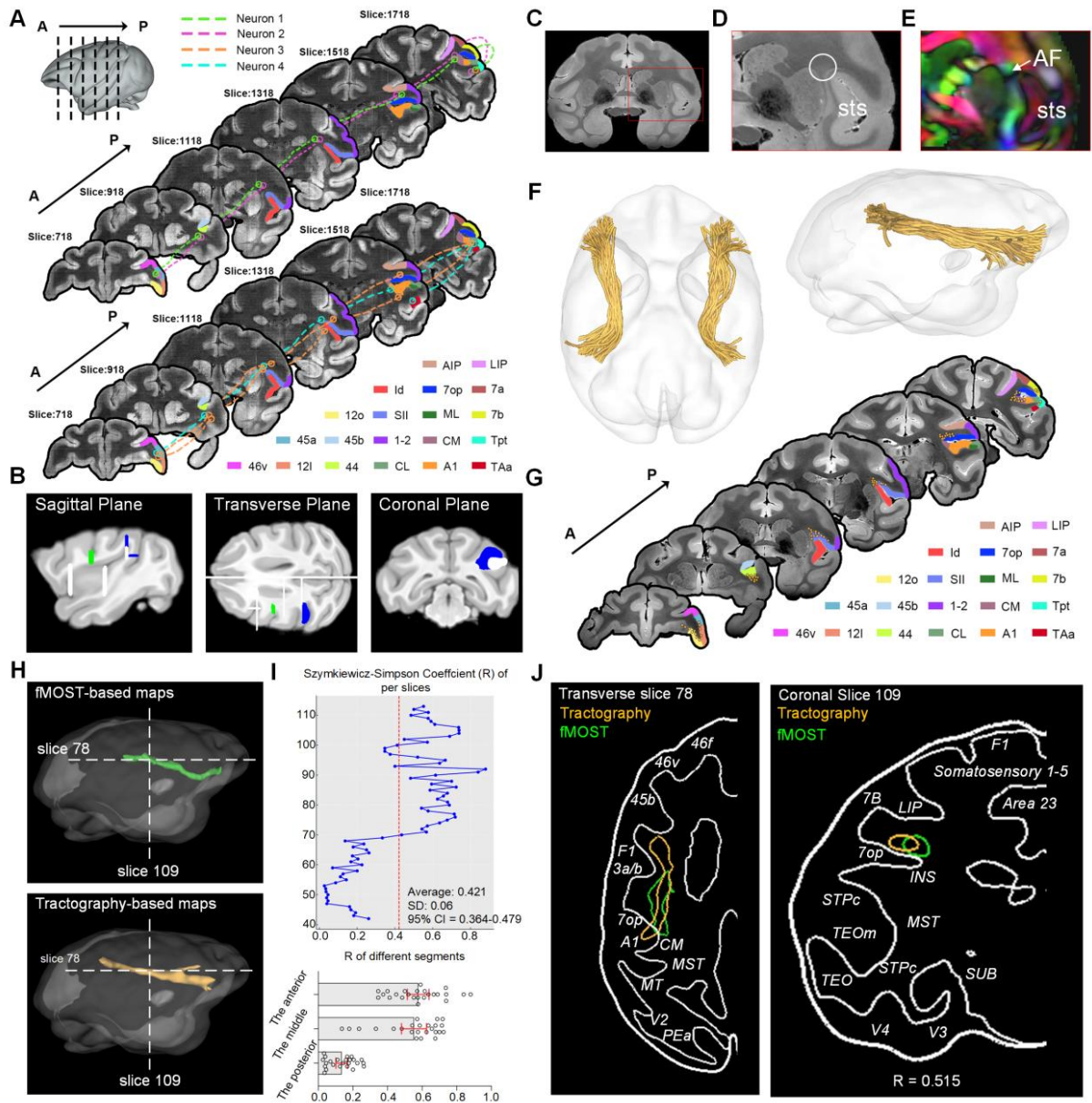

**Figure 4 Cross-scale comparison of the monkey AF pathway by fMOST and dMRI. (A)** In representative fMOST coronal images (rostral to the prefrontal cortex), the trajectories of four reconstructed neurons are depicted alongside adjacent brain regions. **(B)** The refined tractography template for the monkey AF includes exclusion masks (white), seed images (green), and regions of interest (blue). This template was developed by integrating prior templates with neuronal data from fMOST. Two regions of interest were placed dorsally around areas 7op and Tpt. Exclusion masks were positioned at the cingulate sulcus, subcortical gray matter, and lateral sulcus to prevent streamline leakage into the internal capsule and avoid incorrect streamlines crossing between gyri. **(C)** A typical T1-weighted MRI image shown in the coronal plane. **(D)** Magnified view of the red box in **(C)**. The white circled area indicates white matter regions with dense AF neuron projections, set as the seed image as demonstrated

in **(B)**. **(E)** The color-coded fractional anisotropy (FA) map corresponding to **(D)**. **(F)** Tractography result of the AF shown from inferior and lateral perspectives. **(G)** In representative dMRI coronal slices (rostral to the prefrontal cortex), the dMRI-derived AF tract is outlined with small yellow circles on each slice, with adjacent brain regions depicted in different colors. **(H)** The reconstructed course of AF based on fMOST and dMRI tractography, shown in a shared common space. **(I)** The Szymkiewicz-Simpson overlap coefficients between the fMOST- and dMRI-derived AF pathways were calculated, with statistics further segmented into anterior, middle, and posterior sections. **(J)** Specific horizontal and coronal plane slice, marked with dashed lines in **(F)**, showing the overlap of dMRI and fMOST results.

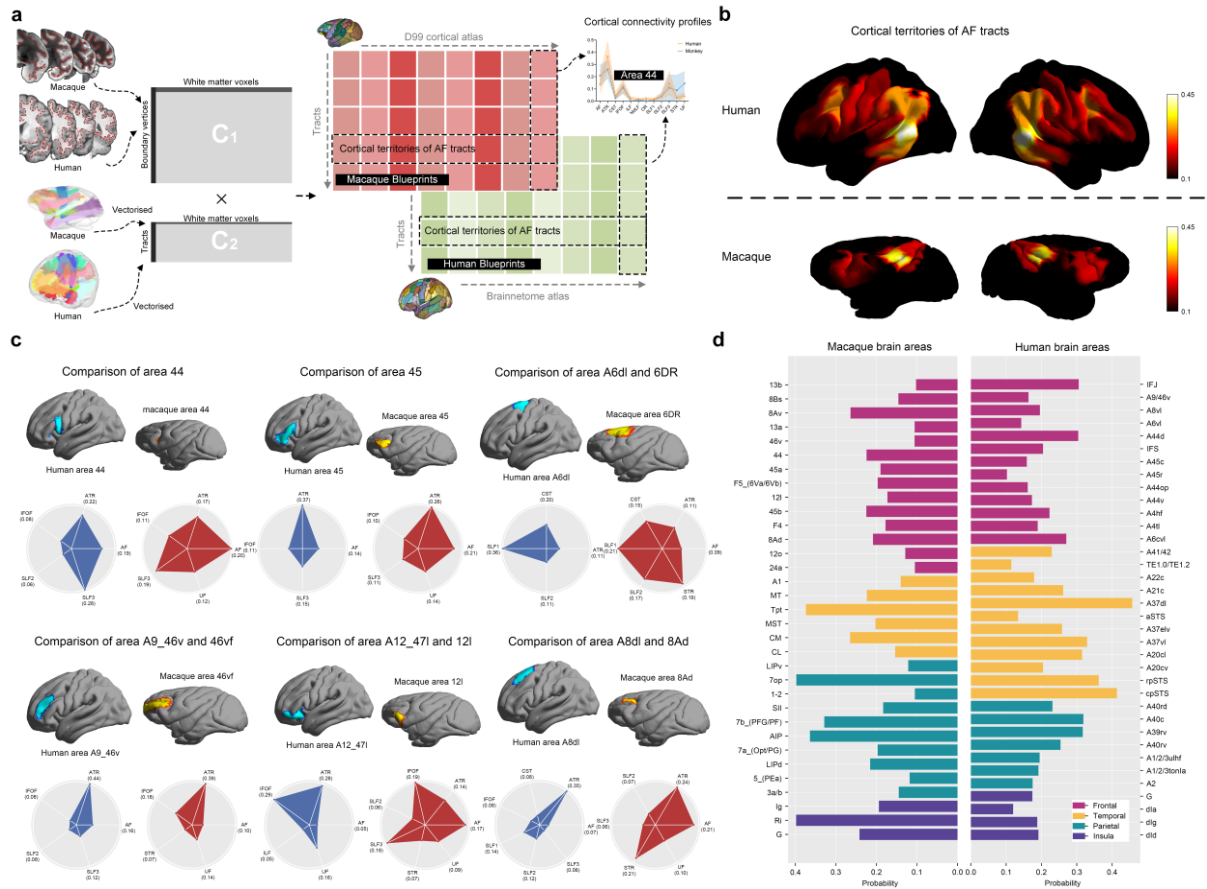

**Figure 5 Characterization of AF connectivity profiles in both species. (A)** A schematic illustrates the process of building connectivity blueprints. Tractography begins at the boundary between white and grey matter, marked by the red outline. The number of visitations across the white matter is calculated to form matrix C1. The tractography data is then vectorized and stacked to generate matrix C2. Finally, matrices C1 and C2 are multiplied to produce the connectivity blueprints for each subject. **(B)** The cortical territories of AF were delineated for both human and macaque brains, corresponding to the columns of the connectivity profile matrix. **(C)** A comparison of connectivity profiles was conducted in homologous frontal regions between human and macaque brains, including areas 44, 45 (45a and 45b), 6DR, 46vf, 12l, and 8Ad. For visualization, only the tracts with a contribution exceeding 0.05 in both species are displayed. **(D)** The connection probabilities of AF to different brain regions are presented. The left side shows AF connection probabilities in different brain areas in monkeys, while the right side shows the corresponding probabilities in various regions of the human brain.

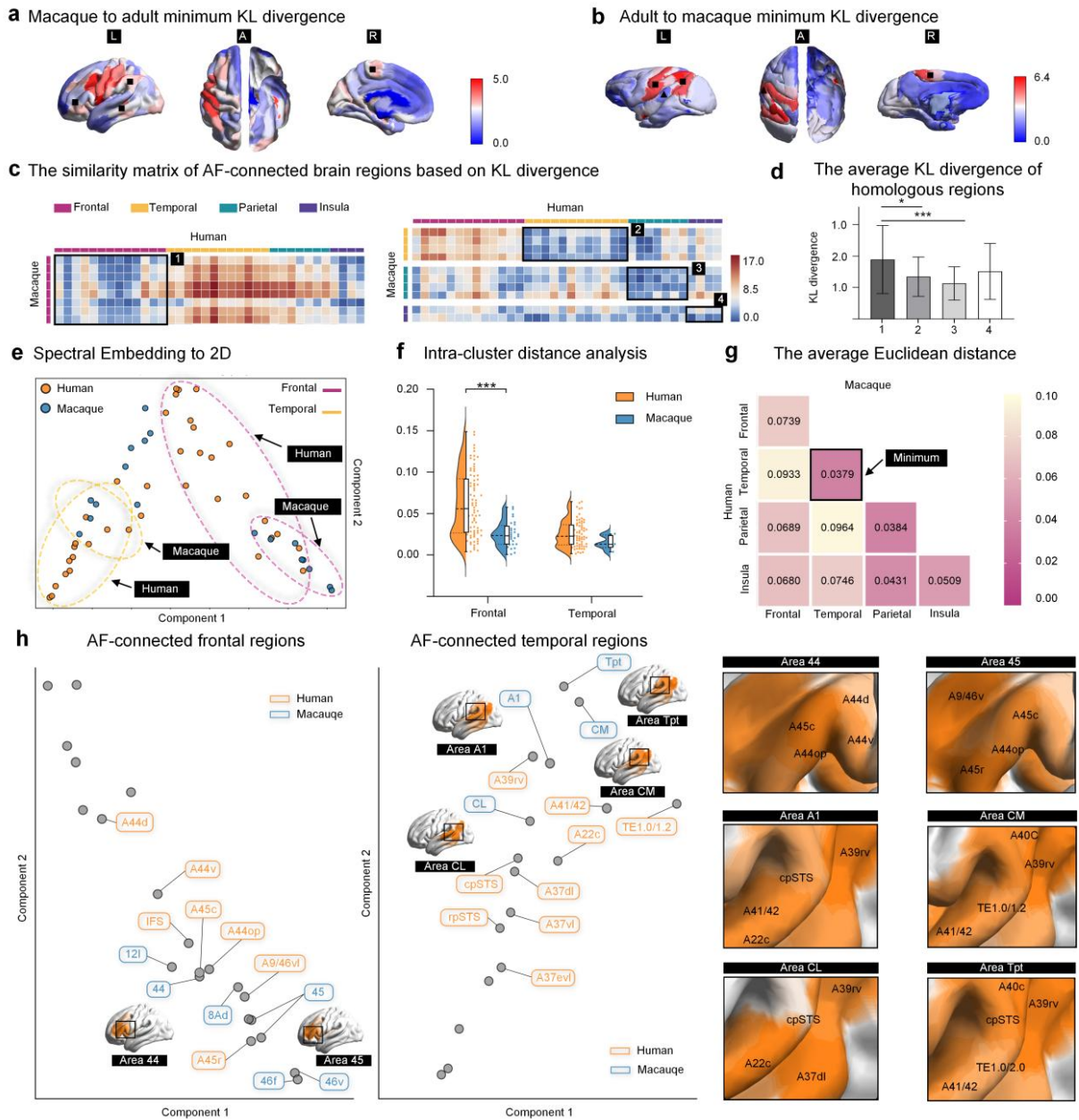

**Figure 6 Divergence of AF connectivity profiles between two species. (A-B)** The minimum Kullback-Leibler (KL) divergence was calculated and visualized for each region, comparing human and macaque brains. Black squares indicate regions that are significantly similar between two species, while black triangles indicate regions that are significantly dissimilar. **(C)** KL divergence in AF-connected regions was computed using cortical parcellation, with the median value taken for all vertices within each pair of regions (details provided in Figure S6). **(D)** The mean KL divergence among regions within the same lobes is indicated by the black box in **(C)**: 1 for frontal, 2 for temporal, 3 for parietal, and 4 for insular regions. **(E)** The exponential of the negative KL divergence matrix was used as a feature set for spectral embedding analysis, with the first two components projected into a 2D space. The similarity

between different parcels was then measured using Euclidean distance (ED). **(F)** Differences in connectivity patterns within the frontal and temporal regions of the two species measured by intra-cluster distances (Mann-Whitney U test). **(G)** In cross-species comparisons of homologous regions, the temporal regions had the highest average ED ($0.074 \pm 0.047$), while the insular regions had the lowest average ED ($0.038 \pm 0.018$). **(H)** AF-connected frontal and temporal regions were analyzed in embedding space, with particular focus on the frontal (e.g., areas 44 and 45) and temporal regions (e.g., Tpt and AF-connected regions of the auditory cortex). Corresponding human brain regions with smaller paired EDs were mapped to visually inspect the (dis-)similarity between specific regions in macaques and their homologous regions in humans. The subplots on the right provide a detailed view of the whole-brain mapping shown on the left.